\documentclass[]{article}
\pdfoutput=1
\usepackage{jheppub}

\usepackage{setspace}
\usepackage{physics}
\usepackage{dsfont}
\usepackage{tensor}
\usepackage[normalem]{ulem}
\usepackage{xcolor}
\usepackage{graphicx}
\usepackage[export]{adjustbox}
\usepackage{mathtools}
\usepackage{mathbbol}
\usepackage[utf8]{inputenc}
\usepackage{amsmath}
\usepackage{bbold}
\usepackage{breqn}
\usepackage{comment}
\usepackage{scalerel,nicefrac}

\usepackage[title]{appendix}

\usepackage[english]{babel}
\usepackage{amssymb}
\usepackage{ragged2e}
\usepackage{etoolbox}
\usepackage{lipsum}
\usepackage{multirow}
\usepackage[most]{tcolorbox}
\usepackage{latexsym}
\usepackage{enumitem}
\usepackage[%
colorlinks=true,%
urlcolor=blue%
]{hyperref}

\usepackage{stackengine,scalerel}
\newcommand\dAlaux{%
  \Shortstack{\rule{12pt}{.6pt}\\
    \rule{.6pt}{10pt}\kern10pt\rule{1.4pt}{10pt}\\
    \rule{12pt}{1.4pt}}%
}
\newcommand\dAl{%
  \setstackgap{S}{0pt}%
  \setstackEOL{\\}%
  \scalerel*{\kern1pt\dAlaux\kern1pt}{\Delta}%
}

\usepackage{titlesec}
\usepackage{url}
\usepackage{caption}
\usepackage{subcaption}
\tcbset{
  highlight math/.style={notitle,nophantom,colframe=red,colback=yellow!25!white}
}

\newcommand*{\colorboxed}{}
\def\colorboxed#1#{%
  \colorboxedAux{#1}%
}
\newcommand*{\colorboxedAux}[3]{%
  \begingroup
    \colorlet{cb@saved}{.}%
    \color#1{#2}%
    \boxed{%
      \color{cb@saved}%
      #3%
    }%
  \endgroup
}

\usepackage{comment}
\usepackage{booktabs} 

\newcommand{\be}{\begin{equation}}
	\newcommand{\ee}{\end{equation}}
\newcommand{\beq}{\begin{equation}}
	\newcommand{\eeq}{\end{equation}}
\newcommand{\bea}{\begin{eqnarray}}
	\newcommand{\eea}{\end{eqnarray}}

\newcommand{\bd}{\partial \mathcal{M}}

\newcommand{\eps}{{\epsilon}}
\newcommand{\apr}{\alpha'}

\title{The GHY boundary term from the string worldsheet to linear order}
   \author[a]{Amr Ahmadain}
   \author[b]{Shoaib Akhtar}
   \author[b]{Rifath Khan}

\affiliation[a] {Department of Physics, Swansea University, Swansea, SA2 8PP, UK}
\affiliation[b] {Stanford Institute for Theoretical Physics, 382 Via Pueblo, Stanford, CA 94305}
\emailAdd{amrahmadain@gmail.com}
\emailAdd{shoaib@stanford.edu}
\emailAdd{rifathkhantheo@gmail.com}

\abstract{Using the method of images we derive the boundary term of the Einstein-$\Gamma^2$ action in half-space from the spherical worldsheet to first order in $\alpha'$ and to linear order in the metric perturbation around flat half-space. The \(\Gamma^2\) action, written down by Einstein more than 100 years ago, includes a boundary term that consists of the Gibbons-Hawking-York action along with two additional terms that are functions of the metric, normal vector, and tangential derivatives. With this boundary term, the total (bulk + boundary) sphere effective action has a well-posed variational principle for Dirichlet boundary conditions. 
}

\begin{document}

\maketitle

\section{Introduction}\label{sec:intro}
It is well known that the bulk Einstein-Hilbert (EH) gravitational action does not have a well-posed variational principle in spacetime manifolds with boundaries. We have to add a boundary term. The structure and coefficient of this boundary term dictate the type of boundary condition we can choose. Adding the Gibbons-Hawking-York \cite{York:1972sj,GH} term, the gravitational action in $D$-dimensional Euclidean manifold $\mathcal{M}$ with boundary $\partial \mathcal{M}$ is
\begin{equation}\label{eq:I_grav}
I_{\text{grav}} = I_{\text{EH}}+I_{\text{GHY}}= -\kappa^2 \int_{\mathcal{M}} d^{D} Y \sqrt{G}\, R -2\kappa^2 \int_{\partial \mathcal{M}} d^{D-1} \xi \sqrt{\gamma}\, K,
\end{equation}
where $\kappa^2 = \frac{1}{16 \pi G_D}$. Here $Y^\mu$ ($\mu \in \{ 1,2, \cdots , D\}$) are coordinates on $\mathcal{M}$ and $\xi^i$ ($i \in \{ 1,2, \cdots , D-1\}$) on $ \partial\mathcal{M}$, $e^{\mu}_i = \partial X^\mu /\partial \xi^i$, $\gamma_{ij}=G_{\mu \nu }e^{\mu}_i e^{\nu}_j$ is the induced metric on $\partial \mathcal{M}$, $K_{ij} = e^{\mu}_i e^{\nu}_j(\nabla_\mu n_\nu+\nabla_\nu n_\mu)/2$ is extrinsic curvature tensor  on $\partial \mathcal{M}$ and $K=\gamma_{ij}K^{ij}$ is the trace of the extrinsic curvature tensor. The variation of $I_{\text{GHY}}$ includes terms that cancel the variations of the normal derivatives of the metric on $\bd$ which arises from the variation of $I_{\text{EH}}$, leaving us with\footnote{We have ignored the boundary total derivative $\nabla_\mu v^\mu$ on $\partial \mathcal{M}$ since it is irrelevant for this work.}
\begin{equation}\label{eq:delta_I_grav}
\delta I_{\text{grav}} = -\kappa^2 \int_{\mathcal{M}} d^{D} Y \sqrt{G}\, \left(R_{\mu\nu} - \frac{1}{2} G_{\mu\nu} R\right)\,\delta G^{\mu\nu} +\kappa^2 \int_{\partial \mathcal{M}} d^{D-1} \xi \sqrt{\gamma}\, (K_{ij}- K\gamma_{ij} )\,\delta \gamma^{ij}.
\end{equation}

In order for $I_{\text{grav}}$ to have well-posed variational principle, the boundary term in \eqref{eq:delta_I_grav} must vanish such that we are left only with Einstein's equation in the bulk for arbitrary variations of the metric $\delta G^{\mu \nu}$. From \eqref{eq:delta_I_grav}, the obvious choice is to fix the metric on the boundary
\begin{equation}\label{eq:DBC}
\left.\delta G^{\mu \nu}\right|_{\partial \mathcal{M}}=0 \iff \delta \gamma^{ij} = 0.
\end{equation}
Therefore, the GHY term (with its coefficient of 2) in \eqref{eq:I_grav} informs us that we should impose Dirichlet boundary conditions to have a well defined variational principle. So, it is the action that guides us to the appropriate boundary condition. 

This also works the other way around - choosing a particular type of boundary condition tells us what the boundary action should be. For example, we can choose to impose \textit{general} Neumann boundary conditions where we fix the momentum conjugate to the induced metric (which is the \textit{densitized} Brown-York boundary stress tensor)\footnote{The Brown-York stress tensor has the form of the undensitized momentum conjugate to the induced metric in the radial (normal) direction on a spacelike surface \cite{Marolf:2012vvz}. Therefore, \eqref{eq:BY_tensor} physically means the codimension-1 $\partial \mathcal{M}$ is free from gravitational energy and flux.} \cite{Brown:1992br} while varying the action, i.e. $\delta T_{ij}\big|_{\partial \mathcal{M}}  = 0$ 
\begin{equation}\label{eq:BY_tensor}
T_{ij}\big|_{\partial \mathcal{M}} \equiv  \frac{2}{\sqrt{\gamma}}\frac{\delta I_{\text{grav}}}{\delta \gamma^{ij}}=  2\kappa^2 (K_{ij}-  K \gamma_{ij})
\end{equation}
However, in this case, the boundary term that must be added in order for the action to have a well-defined variational principle is \textit{different} from the GHY term used to fix $G_{\mu \nu}$ on $\bd$ \cite{Krishnan:2016mcj,Krishnan:2017bte}\footnote{Note that \eqref{eq:Bdy_Neumann} is written in Euclidean signature.}
\begin{equation}\label{eq:Bdy_Neumann}
-(4-D)\, \kappa^2 \int_{\partial \mathcal{M}} \mathrm{d}^{D-1} \xi\sqrt{\gamma}\, K.
\end{equation}  

Now suppose we fix $T_{ij}\big|_{\partial \mathcal{M}}$  in \eqref{eq:BY_tensor} to zero
\begin{equation}\label{eq:BY_Tensor_zero}
T_{ij}\big|_{\partial \mathcal{M}} =0,
\end{equation}
then the variational principle would be trivial - the boundary term in \eqref{eq:delta_I_grav} immediately vanishes and the EH action would, as a result, have a well-defined variational principle on its own\footnote{This is simply because taking the trace of \eqref{eq:BY_Tensor_zero} implies $K=0$.}. No boundary terms are required. This choice leaves the boundary metric \textit{arbitrary} allowing it to adjust dynamically such that it satisfies the boundary equations of motion \eqref{eq:BY_Tensor_zero}\footnote{This is because, alternatively, we can interpret \eqref{eq:BY_Tensor_zero} as a boundary equation of motion.}. 

With this review in mind, we now to proceed to present an overview of our string theory results.

\subsection{Overview of results}
On the string theory side, \cite{Ahmadain_Khan} developed a general approach based on the method of images to extract spacetime boundary terms from the nonlinear sigma model (NLSM) of closed bosonic strings on a spherical worldsheet to first order in $\alpha'$. As a first step, the authors in \cite{Ahmadain_Khan} studied the motion of an dilaton $\Phi(X)$ on a half-space $\mathcal{M} = \mathbb{R}_{+} \times \mathbb{R}^{D-1}$ such that $\bd = \mathbb{R}^{D-1}$ with coordinates $X^\mu=(X^D, \vec{X})$ where $X^D \geq 0$ and $\vec{X}= \{ X^i\}_{i=1}^{D-1}$ represents the $D-1$ tangential directions. $\mathbb{R}_{+}=\mathbb{R} / \mathbb{Z}_2$ $\in [0,\infty)$ is the closed half-line. 


With their worldsheet-based approach, the authors in \cite{Ahmadain_Khan} were able to successfully evaluate the action for $\Phi(X)$ on $\bd$ using Tseytlin's off-shell sphere prescriptions \cite{TSEYTLINMobiusInfinitySubtraction1988,Ahmadain:2022tew} such that the total (bulk+boundary) sphere effective action has a well-posed variational principle
\begin{equation}\label{I_dilaton_action}
I_{\Phi}= -\frac{1}{2}\alpha^{\prime}Z_{\text{nz}}  \left[\int_{\mathcal{M}} d^D Y e^{-2 \Phi} \partial^2 \Phi-  \int_{\partial \mathcal{M}} d^{D-1} Y e^{-2 \Phi} \partial_n \Phi\right].
\end{equation}
    where $\partial_n=-\partial_D$ is the derivative in the direction of the outward-pointing normal and $Z_{\text{nz}}$ is the free sphere partition function. Note that we have chosen the wall to be at $Y^D = 0$ and taken the boundary coordinates to be $\xi^i := Y^i$. Using the equation of the motion for $\Phi$, \cite{Ahmadain_Khan} wrote down the classical on-shell action for the dilaton as a sum of two boundary contributions that each provided half the action \cite{KT:2001,chen2023black}
\begin{equation}\label{I_dilaton_onshell}
I_{\Phi,\partial \mathcal{M}}= \alpha^{\prime} Z_{\text{nz}}\int_{\partial \mathcal{M}} d^{D-1} Y e^{-2 \Phi} \partial_n \Phi.
\end{equation}
Here, $\partial_n \Phi$ is the normal derivative of dilaton at $\bd$. 

This work is a generalization of \cite{Ahmadain_Khan} to curved target spaces with a metric $G_{\mu \nu}(Y)$ expressed as an arbitrary small perturbation $h_{\mu \nu}$ around half-space with a flat Euclidean metric. In this case, $\mathcal{M}$ and $\partial \mathcal{M}$ are fully dynamical while $\Phi(Y)$ is kept fixed, in contrast to \cite{Ahmadain_Khan}. We otherwise follow the same setup and conventions defined above.

In this paper, we provide a string worldsheet derivation of the boundary term in the Einstein-$\Gamma^2$ action\footnote{The action density of the Einstein-$\Gamma^2$ \cite{Dyer:2008hb} is $L_{\Gamma^2} = -G^{\alpha \beta}(\Gamma^\nu_{\mu \alpha} \Gamma^{\mu}_{\nu \beta} - \Gamma^{\mu}_{\mu \nu} \Gamma^{\nu}_{\alpha \beta}) = -R - \frac{1}{\sqrt{G}}\partial_\mu \left(\sqrt{G}\, A^\mu\right)$  with $A^{\mu} =  -G^{\alpha \beta} \Gamma^{\mu}_{\alpha \beta} + G^{\mu \alpha} \Gamma^{\beta}_{\alpha \beta}$,
which in terms of $K$ can be expressed as $n_\mu A^{\mu}= 
2 K - 2 \gamma^{\alpha \beta} \partial_\alpha n_\beta + n^\beta \gamma^{\alpha \mu} \partial_\alpha G_{\mu \beta}$.} \cite{Einstein:1916cd} to $O(\apr)$ (and to linear order in $h_{\mu \nu}$) using the method of images. The Einstein-\(\Gamma^2\) action includes a boundary term that consists of the GHY action along with two additional terms \cite{Einstein:1916cd,Dyer:2008hb,Padmanabhan_2010,Parattu:2016trq,Chakraborty:2016yna} 
\begin{align}\label{eq:I_bdy_intro}
I_{\Gamma^2}=-\kappa^2 \int_{\mathcal{M}} d^{D} Y \sqrt{G}\, R-  \kappa^2\int_{\partial \mathcal{M}} \mathrm{d}^{D-1} Y \sqrt{\gamma} \left(2 K - 2 \gamma^{\alpha \beta} \partial_\alpha n_\beta + n^\beta \gamma^{\alpha \mu} \partial_\alpha G_{\mu \beta}\right).
\end{align}
For Dirichlet boundary conditions \eqref{eq:DBC}, \(2K\) is not the only viable option for the boundary term, as we can always add any function of the metric, normal vector, and tangential derivatives \cite{Dyer:2008hb,Padmanabhan_2010,Parattu:2016trq}. The difference from the GHY term in \eqref{eq:I_bdy_intro} serves as an example of such a function of the boundary data. The variation of this function vanishes when we use \eqref{eq:DBC} or \eqref{eq:BY_Tensor_zero} and we still end up with Einstein's equations of motion and thus, a well-posed variational principle.\footnote{As pointed out in chapter 6 of \cite{Padmanabhan_2010}, the Einstein boundary and GHY terms are equal only when the shift function is fixed on $\bd$ but their variations are equal however, if we fix the whole metric is fixed on $\bd$.} 

The total classical (tree-level) effective action for strings in a half-space we derive including the boundary term for the dilaton \eqref{I_dilaton_onshell} is given by the sum of the bulk and boundary contributions after applying Tsyetlin's off-shell sphere prescriptions to eliminate the logarithmic divergences from the respective partition functions
\begin{align}\label{eq:I_total}
I_{\text{sphere,HS}} &= -\frac{\partial}{\partial \ln \eps} \scalerel*{(}{\strut}Z_{\text{bulk}} + Z_{\text{wall}}\scalerel{)}{\strut} \\ \nonumber
&= -\frac{1}{4}\alpha' Z_{\text{nz}} \int_{\mathcal{M}} \mathrm{d}^D Y\, \sqrt{G}\, e^{-2 \Phi} (R + 2\nabla^2 \Phi) \\ \nonumber 
& \quad - \frac{1}{4}\alpha' Z_{\text{nz}}  \int_{\partial \mathcal{M}} \mathrm{d}^{D-1} Y \sqrt{\gamma}\, e^{-2 \Phi} \left(2 K - 2 \partial_n \Phi + n^\beta \gamma^{\alpha \mu} \partial_\alpha G_{\mu \beta}\right).
\end{align}
Note that \eqref{eq:I_total} is only proportional to \eqref{eq:I_bdy_intro} since we can't fix the gravitational coupling constant in the off-shell NLSM approach.

Upon varying \eqref{eq:I_total} and before imposing the dilaton equation of motion, the action \eqref{eq:I_total} exhibits a well-defined variational principle that is compatible with the Dirichlet boundary conditions \eqref{eq:DBC} \textit{or} special Neumann boundary conditions \eqref{eq:BY_Tensor_zero}, but \textit{not} with \eqref{eq:BY_tensor}.\footnote{Imposing special Neumann boundary conditions \eqref{eq:BY_Tensor_zero} on the two additional terms in \eqref{eq:I_total} will enforce a gauge fixing condition on the normal vector $n^{\alpha}$ and the induced metric $\gamma^{\alpha \beta}$ causing them both to vanish. One simple gauge choice is Gaussian normal coordinates where $n_\mu = (-1,0, 0, \dots, 0)$ and the lapse function $N^{\alpha}=0$.}

Tseytlin derived $\int R$ in \cite{Tseytlin:1990mv,TseytlinZeroMode1989}, but only up to some total derivative terms, which were dropped because they yield boundary terms via integration by parts. This was acceptable in his case because he was considering manifolds with no boundary or asymptotically flat spacetimes where those boundary terms would vanish. The omission of these total derivative terms was noted and were written down in \cite{Ahmadain:2022eso}. However, if we now consider manifolds with boundaries, these terms cannot be ignored. In \cite{Ahmadain:2024uom}, these total derivative terms were converted into boundary terms using integration by parts for a constant $\Phi$ to get
\begin{equation}
\label{eq:I_TD_intro}
I_{\text{bulk}}=-\frac{1}{4}\alpha^{\prime} Z_{\text{nz}} \left[\int_{ \mathcal{M}}\mathrm{d}^D Y\, \sqrt{G} e^{-2\Phi} R + \int_{\partial \mathcal{M}} \mathrm{d}^{D-1} Y \sqrt{\gamma}\, e^{-2\Phi}\left(K-\gamma^{\alpha \beta} \partial_\alpha n_\beta-\frac{1}{2} n^\mu n^\alpha n^\beta \partial_\mu G_{\alpha \beta}\right)\right],
\end{equation}
which we also derive independently at linear order in section \ref{sec:I_total}.

The above action \eqref{eq:I_TD_intro} clearly does not have a well-defined variational principle. However, when we take the presence of the wall into account and compute the half-space sphere partition function $Z_{\text{HS}}$, we find an additional boundary term originating from the wall itself
\begin{equation}\label{eq:I_wall_intro}
I_{\text{wall}} = -\frac{1}{4}\apr Z_{\text{nz}}\int_{\partial \mathcal{M}} \mathrm{d}^{D-1} Y \sqrt{\gamma} \,e^{-2\Phi}\left(K + n^\beta \gamma^{\alpha \mu} \partial_\alpha G_{\mu \beta} + \frac{1}{2} n^\alpha n^\beta n^\mu \partial_\alpha G_{\mu \beta}\right).
\end{equation}
We derive \eqref{eq:I_wall_intro} to linear order in the metric perturbation using the method of images in this paper. We emphasize that it is only when we take the wall term \eqref{eq:I_wall_intro} into account that we get the total sphere action \eqref{eq:I_total}, which now has a well-defined variational principle for Dirichlet boundary conditions. This is the main result of our work.

Notice that choosing our wall to be fixed at $Y^D=0$ automatically gauge fixes the normal 1-form to $n_\mu=(-1,0,0, \ldots,0)$, which causes the term $\gamma^{\alpha \beta} \partial_\alpha n_\beta$ in \eqref{eq:I_bdy_intro} to vanish. However, this choice leaves the shift vector $N^\alpha$ unfixed. We will explain this in more detail in appendix \ref{app:gauge_terms}. Observe that $\gamma^{\alpha \beta} \partial_\alpha n_\beta$ appears in \eqref{eq:I_TD_intro} but for consistency, we will also set it to zero by fixing $n_\mu$ to be constant. 


Taking into account the boundary term for the dilaton \eqref{I_dilaton_onshell}, the classical \textit{on-shell} action can now be expressed as the normal derivative of the generalized volume $(e^{-2\Phi} \sqrt{\gamma}$) \cite{KT:2001,chen2023black}\footnote{We have omitted the gauge-dependent terms in \eqref{eq:I_onshell} since they vanish while varying the action.}
\begin{equation}\label{eq:I_onshell}
I_{\partial \mathcal{M}}=-\frac{1}{4}\alpha^{\prime} Z_{\text{nz}} \int_{\partial \mathcal{M}} \mathrm{d}^{D-1} Y \sqrt{\gamma} e^{-2\Phi}\left(2K - 4\partial_n\Phi \right) =- \frac{1}{2}\alpha^{\prime} Z_{\text{nz}}\int_{\partial \mathcal{M}} \mathrm{d}^{D-1} Y \partial_n (e^{-2\Phi} \sqrt{\gamma}).
\end{equation}
Observe that the action $I_{\partial \mathcal{M}}$ is on-shell only as a consequence of using the dilaton equation of motion. When evaluated on a string solution  e.g. the cigar \cite{KT:2001,Chen-LargeD:2021,chen2023black}, it immediately yields the black hole entropy. Generally speaking, knowing how to directly extract target space boundary terms from the worldsheet is fundamentally important for a microscopic understanding of black hole microstates \cite{SU-1994,Ahmadain:2022eso}. In this work, however, our worldsheet derivation of \eqref{eq:I_total} is off-shell; we deform about half-space but note that the target space action \eqref{eq:I_total} itself is locally background-independent. 

\subsection{Layout of paper}
In section \ref{sec:MOI}, we discuss the general concept of the method of images and demonstrate how it is applied to the string worldsheet to derive boundary terms in target space. Section \ref{sec:Setup} provides the details of the worldsheet deformation used in computing the wall term $I_{\text {wall}}$. In section \ref{sec:EinsteinBoundaryAction}, we put things together. We first evaluate the partition function for strings in the half-space and extract \(I_{\text{wall}}\) by using Tseytlin's off-shell sphere prescriptions. We then we add it to the boundary term from the bulk tree-level action \eqref{eq:I_TD_intro} to obtain the Einstein boundary term \eqref{eq:I_bdy_intro} in the gauge where the normal 1-form is constant. We show how our calculations are able to capture the relative factor of ``2" between the EH and GHY actions in \eqref{eq:I_total}. Finally in section \ref{sec:discussion}, we present a subtlety related to imposing boundary conditions on the string at the wall and end with some potential future directions.




\section{The method of images on the worldsheet}\label{sec:MOI}
In this section, we outline the general idea behind the method of images and how we use it on the string worldsheet to derive boundary terms in target space. To keep the presentation clear, we omit some but not all of the technical details that may obscure the main idea and instead present them in section \ref{sec:Setup}. The content of this section is a generalized summary of sections 2, 3 and appendix A in \cite{Ahmadain_Khan}.

The method of images is at its core a calculational (doubling) trick that allows us to evaluate a partition function in a doubled space with the reflected field $\Psi_{\text{ref}}(X)$ such that the half-space partition function $Z_{\text{HS}}$ can be expressed in terms of its whole space counterpart $Z_{\text{RS}}$
\begin{align}\label{eq:HL_action}
Z_{\text{HS}} &= \int_{X^D \geq 0} [DX] e^{-I_{\text{w.s.}}[X,\Psi]} =\frac{1}{2}\, Z_{\text{RS}} = \frac{1}{2}\int [DX] e^{-I_{\text{w.s.}}[X,\Psi_{\text{ref}}]}.
\end{align}
Here, $Z_{\text{HS}}$ represents the path integral of spherical worldsheet embeddings on the right side of the wall \((X^D \geq 0)\). Generally speaking, the method of images transforms a boundary value problem into a problem without boundaries and therefore, eliminates the need to explicitly handle boundary terms in calculations with explicit delta functions or Lagrange multipliers. For the \textit{worldline} \cite{Bastianelli:2006hq,Bastianelli:2008vh,Bastianelli:2009mw}, it is the sign we use in the \textit{spacetime} heat kernel when the particle bounces off the wall that determines the boundary condition. Imposing a Dirichlet boundary condition comes with a $(-)$ sign \cite{Clark1980half,Asorey:2007zza}. Neumann boundary conditions \eqref{eq:BY_Tensor_zero}, on the other hand, come with a $(+)$ sign in \eqref{eq:HL_action}. Extending this to the string worldsheet presents us with a subtlety that we discuss in section \ref{sec:discussion}.

Let $\Psi(X)$ be \textit{any} massless\footnote{Although we can in principle also include the closed string tachyon.} closed or open bosonic string mode, scalar or tensor and $\Psi_{\text{sol}}(X)$ denotes field configurations that solves the string spacetime equations of motion, i.e. $\Psi_{\text{sol}}(X)$ is an on-shell string background. The background $\Psi_{\text{sol}}(X)$ is then deformed by $\psi(X)  \ll 1$ (which is defined only for $X^D \geq 0$) 
\begin{equation}
    \Psi(X) = \Psi_{\text{sol}}(X)+\psi(X).
\end{equation}
Now in the reflected space we extend the field as follows (this is now defined for all $X^D \in \mathbb{R}$)
\begin{equation}\label{eq:Wall_deformation}
\Psi_{\text{ref}}(X)=\Psi_{\text{ref}}^{\text{sol}}(X)+\psi_{\text{ref}}(X),
\end{equation}
where
\begin{equation}
    \begin{aligned}
\Psi_{\text{ref}}^{\text{sol}}(X) & = \Psi_{\text{sol}}(X)\Theta(X^D)+\mathcal{R}\circ\Psi_{\text{sol}}(X)\Theta(-X^D) \\
   \psi_{\text{ref}}(X) & = \psi(X)\Theta(X^D)+\mathcal{R}\circ\psi(X)\Theta(-X^D).
\end{aligned}
\end{equation}
Here $\mathcal{R}\circ \Psi$ is the reflection of $\Psi$ under $(X^D,\vec{X})\to (-X^D, \vec{X})$. 

Our main goal now is to understand how to systematically calculate $Z_{\text{RS}}$ with the deformation \eqref{eq:Wall_deformation}. As in any NLSM expansion, we split the string $X^\mu$ into a zero mode $Y^\mu$ and a non-zero fluctuating mode $\eta^\mu(z)$

\begin{equation}
    X^\mu(z) = Y^\mu + \sqrt{\apr} \eta^\mu(z),
\end{equation}
and express the NLSM action as the action $I_{\text{sol}}$ of the on-shell saddle $\Psi_{\text{sol}}$ perturbed by $\Hat{V}_{\text{def}}$ 
\begin{align}\label{eq:WS_deformation}
I_{\text{w.s.}}&=I_{\text{sol}}[\eta,g,\Psi_{\text{ref}}^{\text{sol}}] + \Hat{V}_{\text{def}}[\eta ,g, \psi_{\text{ref}}],
\end{align}
where the deformation $\Hat{V}_{\text{def}}$  in terms of \eqref{eq:WS_deformation} is given by
\begin{equation}
\Hat{V}_{\text{def}} \coloneqq \frac{1}{4 \pi } \int d^2 z \sqrt{g} \, P(\eta(z)) \psi_{\text{ref}}(Y+\sqrt{\apr}\eta(z)).
\end{equation}
Here $g$ is the determinant of the the worldsheet metric  $g_{ab}(z)$ and $P(\eta(z))$ is a polynomial in the derivatives of $\eta(z)$. 


The one-point function $\langle \Hat{V}_{\text{def}}\rangle$ is defined by 
\begin{align}\label{eq:V_def}
\langle \Hat{V}_{\text{def}}\rangle &=\frac{1}{Z_{\text{nz}}} \int [D\sqrt{\apr}\eta] e^{-I_{\text{sol}}[\eta,g,\Psi_{\text{ref}}^{\text{sol}}]} \,\Hat{V}_{\text{def}} \\ \nonumber
&= \frac{1}{4 \pi } \int d^2 z \sqrt{g}  \left\langle P(\eta(z)) \psi_{\text{ref}}(Y+ \sqrt{\apr}\eta(z))\right\rangle.
\end{align}
Here $Z_{\text{nz}}$ is the path integral for the non-zero modes (the free sphere partition function).
$Z_{\text{HS}}$ in \eqref{eq:HL_action} is obtained by integrating\footnote{In this paper, we calculate boundary terms only up to linear order in the deformation \(\psi\), neglecting contributions from second terms in \eqref{eq:Z_HL}. However, for the dilaton studied in \cite{Ahmadain_Khan}, it is necessary to consider the second order contributions because the action density involves second order derivatives. Integration by parts yields linear order terms up to an exponential factor \(e^{-2\Phi} \partial^2  \Phi\). But, it can be explicitly verified that these terms all contain power-law divergences, which can be thus removed by local counterterms.} $\langle \Hat{V}_{\text{def}}\rangle$  over the zero mode $Y$
\begin{align}\label{eq:Z_HL}
Z_{\text{HS}}&=\frac{1}{2} Z_{\text{nz}} \int d^D Y  \left(1-\langle \Hat{V}_{\text{def}} \rangle+ \frac{1}{2}\langle \Hat{V}^2_{\text{def}} \rangle+ \langle O(\Hat{V}^3_{\text{def}}) \rangle \right),
\end{align}
 Evaluating \eqref{eq:V_def} in regions of the target space both far from and near the wall is key to understanding how the method of images works. To clarify this, let us Taylor expand $\langle \Hat{V}_{\text{def}} \rangle$ about the point $Y$ far from the wall\footnote{We have also taken $P(\eta)=1$ in \eqref{eq:one-point-function-expansion} to make the presentation simple. Details of the calculation for a perturbation by a tensor field, where $P(\eta)$ is not zero, will be presented in sections \ref{sec:Setup} and \ref{sec:EinsteinBoundaryAction}. }, where $|Y^D| \gg \sqrt{\alpha'}$,
\begin{align}\label{eq:one-point-function-expansion}
\left\langle \psi_{\text{ref}}\left(Y+\sqrt{\apr}\eta(z)\right)\right\rangle &= \psi\left(Y\right) + \sqrt{\apr}\partial_\mu \psi\left(Y\right) \left\langle \eta^\mu (z) \right\rangle  + \frac{\apr}{2} \partial_\mu \partial_\nu \psi\left(Y\right) \left\langle \eta^\mu(z) \eta^\nu (z) \right\rangle + \cdots.
\end{align}
In this case, the string's motion is unrestricted, or effectively free, as it does not detect the presence of the boundary. Mathematically, the action of the free field $\eta(z)$ is quadratic (Gaussian), leading to $\langle \eta^\mu(z) \rangle = 0$.

\textit{On the other hand, the motion of the string in the close vicinity of the wall where $\abs{Y^D}\sim \sqrt{\apr}$ is quite different. In this regime, the singular nature of the deformation in \eqref{eq:Wall_deformation} becomes important, allowing us to extract the physics of the string close to the target space boundary.} Specifically, when $Y$ is near the wall, the kink in $\psi_{\text{ref}}$ does matter, and to linear order, $\psi_{\text{ref}}$ can be approximated as
\begin{align}\label{eq:psi_def}
{\psi_{\text{ref}}}_\bullet(X^D,\vec{X}) \approx (\text{sgn} (X^D))^{p(\bullet)}\psi_{\bullet}(0,\vec{X}) + (\text{sgn} (X^D))^{p(\bullet)}\partial_D \psi_\bullet(0,\vec{X}) \abs{X^D}\,,
\end{align}
where 
\begin{equation}\label{eq:derivative_at_wall}
\partial_D \psi(0,\vec{X}) := \lim_{X^D \to 0^+} \partial_D \psi(X^D,\vec{X}),
\end{equation}
More generally
\begin{align}
{\psi_{\text{ref}}}_\bullet(X^D,\vec{X}) = (\text{sgn} (X^D))^{p(\bullet)}\sum_k \frac{1}{k!}\partial^k_D {\psi_{\text{ref}}}_\bullet(0,\vec{X})|X^D|^k
\end{align}

Here $\bullet$ represents components of the (possibly tensor) field $\psi(X^D,\vec{X})$ in some representation and $p(\bullet)$ refers to their reflection parity. $p(\bullet)=0$ means the component $\bullet$ does \textit{not} change sign across the wall and thus develops a kink after reflection. For these components, $\partial_D {\psi_{\text{ref}}}_\bullet$ contains a $\delta(X^D)$-localized interaction on the boundary. In contrast to the region far from the boundary, this kink gives rise to a non-zero one-point function. On the other hand, components with $p(\bullet)=1$ do change sign upon reflection, do not develop a kink and hence give no boundary contribution. Although $p(\bullet)=1$ components do develop a discontinuity, it turns out they don't contribute to the partition function as explained below.



Taking the expectation value of \eqref{eq:psi_def} to $O(\sqrt{\apr})$ gives
\begin{equation}\label{eq:one_point_function_psi}
\begin{aligned}
\left\langle{\psi_{\text{ref}}}_\bullet\left(Y+\sqrt{\apr}\eta(z)\right)\right\rangle &= \left[\text{erf}\left(\frac{Y^D}{\sqrt{\alpha' \Omega} }\right)\right]^{p(\bullet)} \psi_\bullet( 0,\vec{Y})+\partial_D \psi_\bullet( 0,\vec{Y})\left( u(Y^D) \delta_{p(\bullet),0}+Y^D \delta_{p(\bullet),1}\right),
\end{aligned} 
\end{equation}
where
\begin{equation}\label{eq:u(Y^D)}
u\left(Y^D\right)\coloneqq \left\langle\abs{Y^D+\sqrt{\apr}\eta^D(z)}\right\rangle.
\end{equation}
To get to this we have used the main result derived in Appendix A of \cite{Ahmadain_Khan} to calculate
\begin{equation}
     \left \langle \text{sgn} (X^D) \right\rangle = \text{erf}\left(\frac{Y^D}{\sqrt{\alpha' \Omega} }\right).
\end{equation}
Later when we integrate over the near wall region $\int_{|Y^D| \sim \sqrt{\alpha'}}dY^D (\cdots ) $, the $\text{erf}\left(\frac{Y^D}{\sqrt{\alpha' \Omega} }\right)$ term (which arises for $p(\bullet)=1$ components) will not contribute simply because it is an odd function. This integral also gives a factor of $\sqrt{\alpha'}$ and  makes \eqref{eq:one_point_function_psi} contribute at O($\alpha'$). The calculation of $u(Y^D)$ represents the backbone of this work and of \cite{Ahmadain_Khan}. Below, we comment on its structure and meaning.

Local vertex operators in string theory are typically expressed as worldsheet integrals of polynomials involving the derivatives of $X(z)$. Singular operators such as $|Y^D+\sqrt{\apr}\eta^D(z)|$ do not belong to this class of operators. However, the heat kernel regulator used in the evaluation of $u(Y^D)$ effectively smooths out these short-distance divergences and thus, regularizes the expression.

In Appendix A of \cite{Ahmadain_Khan}, the expectation value\footnote{In a compact CFT, the expectation value of primary operators due to conformal symmetry. Please see \cite{Kraus:noncompactCFT:2002} for why one-point functions of nonrenormalizable operators do not vanish in a noncompact CFT.}  $\left\langle\abs{\eta^D\left(z\right)}^N\right\rangle$ was calculated as
\begin{align}\label{eq:one-point-eta}
\left\langle\abs{\eta^D\left(z\right)}^N\right\rangle&=\frac{1}{Z_{\mathrm{nz}}^{(D)}} \int\left[D \sqrt{\alpha^{\prime}} \eta^D\right] e^{-I_{\text{P}}^{(D)}[\Psi_{\text{P}},\eta^D,g]} \abs{\eta^D\left(z\right)}^N \\ 
&=\frac{1}{Z_{\mathrm{nz}}^{(D)}} \int\left[D \sqrt{\alpha^{\prime}} \eta^D\right] e^{-I_{\text{P}}^{(D)}[\Psi_{\text{P}},\eta^D,g]} \abs{\eta^D\left(z\right)}^N \int_{-\infty}^{\infty} d \xi \, \delta\left(\xi-\eta^D(z)\right)\\
&=\frac{1}{Z_{\mathrm{nz}}^{(D)}} \int_{-\infty}^{\infty} d \xi~ \abs{\xi}^N\int\left[D \sqrt{\alpha^{\prime}} \eta^D\right] e^{-I_{\text{P}}^{(D)}[\Psi_{\text{P}},\eta^D,g]}   \, \delta\left(\xi-\eta^D(z)\right),
\end{align}
where we have inserted a Dirac delta function in the second line and mathematically manipulated the expression to replace the operator $|\eta^D(z)|^N$ by $\xi$ and pull it out of the integral after changing the order of integration. This way we only have to calculate the partition function with a simpler constraint of fixing $\eta^D$ at the point $z$ to be $\xi$. And finally we get
\begin{equation}
\label{eq:eta^N}
\begin{aligned}
    \left\langle\abs{\eta^D\left(z\right)}^N\right\rangle&= \frac{1}{\sqrt{\pi}} \Gamma \left( \frac{1+N}{2}\right)\left( \sum_{n=1}^\infty \frac{2n+1}{n(n+1)}e^{-\epsilon n (n+1)}\right)^{N/2}\\
    &= \frac{1}{\sqrt{\pi}} \Gamma \left( \frac{1+N}{2}\right)\left(-\ln \epsilon\right)^{N/2}+\cdots
\end{aligned}
\end{equation}

To regularize the ultraviolet (UV) divergence in \eqref{eq:one-point-eta}, we inserted the heat kernel factor $e^{-\epsilon \nabla^2}$ into $I^{(D)}_{\text{P}}$ where $\epsilon$ is the UV cutoff on the worldsheet. Power-law divergences are removed by local counterterms while logarithmic divergences are universal and are removed by Tseytlin's sphere prescriptions. (See \cite{Ahmadain:2022eso} for more details).


The quantity $u(Y^D)$ \eqref{eq:u(Y^D)} can be understood intuitively. As noted in section 5 of \cite{Ahmadain_Khan}, $u(Y^D)$ represents the average absolute distance (or length in Euclidean target space) of the string near the wall, corresponding to the first moment of the half-normal distribution. In the context of reflected Brownian motion in stochastic systems \cite{morters2010brownian,karatzas2014brownian}, $u(Y^D)$ is closely related to the local boundary time \cite{it1965diffusion,levy1965processus} which quantifies the amount of time (or length) the string spends near the boundary,  effectively counting the number of times the string hits and reflects back from the wall. 

It also turns out $u(Y^D)$ is the solution to the one-dimensional heat equation with a non-homogeneous source term \cite{Ahmadain_Khan}
\begin{equation}
\frac{d^2 u}{d\left(Y^D\right)^2}=\frac{2}{\sqrt{\alpha^{\prime} \pi \Omega}} e^{-\frac{\left(Y^D\right)^2}{\alpha^{\prime} \Omega}} .
\end{equation}
where the explicit evaluation of $u(Y^D)$ gives
\begin{align}\label{eq:u(Y)_sol}
u(Y^D) & =\sqrt{\Omega} \frac{\sqrt{\alpha^{\prime}}}{\sqrt{\pi}} e^{-\frac{\left(Y^D\right)^2}{\alpha^{\prime} \Omega}}+Y^D \operatorname{erf}\left(\frac{Y^D}{\sqrt{\alpha^{\prime} \Omega}}\right) \\
& =\int_0^{Y^D} d Y^D \operatorname{erf}\left(\frac{Y^D}{\sqrt{\alpha^{\prime} \Omega}}\right)+ C .
\end{align}
With the following boundary conditions
\begin{equation}
u'(0) = \operatorname{erf}\left(\frac{0}{\sqrt{\alpha' \omega}}\right) = 0, \quad u'(\infty) = \operatorname{erf}\left(\frac{\infty}{\sqrt{\alpha' \omega}}\right) = 1, \quad u(0) = \sqrt{\Omega} \frac{\sqrt{\alpha^{\prime}}}{\sqrt{\pi}} = \sqrt{\alpha^{\prime}}\left\langle\left|\eta^D\left(z_0\right)\right|\right\rangle,
\end{equation}
the constant is $C = u(0)$, which is the value of $u(Y^D)$ at $Y^D=0$, as found in \cite{Ahmadain_Khan}. Notice also that the structure of \eqref{eq:u(Y)_sol} shows that $u(Y^D)$ is the integral of a total derivative, which corresponds to a boundary term. Integrating it over the half-line yields its length, resulting in an infrared divergence. The details of how we regularize this divergence are discussed in Chapter \ref{sec:EinsteinBoundaryAction}.


\section{The metric deformation of the worldsheet theory}\label{sec:Setup}
In this section, we present the details of the worldsheet NLSM deformation that we use in computing the wall term \eqref{eq:I_wall_intro}. We will closely follow the steps outlined in section \ref{sec:MOI}.

While the deformation \eqref{eq:Wall_deformation} in the previous section was about general string solutions, in this paper we deform about Euclidean flat half-space 
\begin{equation}
    G_{\mu \nu}(X) = \delta_{\mu \nu}+h_{\mu \nu}(X)~~\forall~X^D \geq 0
\end{equation}
Then we go to the reflected space via the method of images where the metric is 
${h}^{\text{ref}}_{\mu\nu}(X)$
\begin{equation}
{G}^{\text{ref}}_{\mu\nu}(X)=\delta_{\mu\nu} +{h}^{\text{ref}}_{\mu\nu}(X),
\end{equation}
where
\begin{equation}
{h}^{\text{ref}}_{\mu\nu}(X) = h_{\mu\nu}(X^D,\vec{X})\Theta(X^D)+(\mathcal{R}\circ h)_{\mu\nu}(X^D,\vec{X})\Theta(-X^D),
\end{equation}
and the reflected the rank-2 tensor $(\mathcal{R}\circ h)_{\mu\nu}(X^D,\vec{X})$ is given by (indices are not contracted)
\begin{equation}
(\mathcal{R}\circ h)_{\mu\nu}(X^D,\vec{X}) = (-1)^{\delta_{\mu,D}}(-1)^{\delta_{\nu,D}} h_{\mu \nu}(X^i,-X^D).
\end{equation}
Notice that the $h_{iD}$ element flips sign under reflection.

Following the steps in section \ref{sec:MOI}, the one-point function of the metric deformation $\Hat{V}_{\text{def}}$ is given by
\begin{equation}\label{eq:EV_V_h}
  \langle  \Hat{V}_{\text{def}} \rangle  = \frac{1}{4\pi} \int d^2 z \sqrt{g}  g^{ab}  \langle \partial_a \eta^\mu \partial_b \eta^\nu {h}^{\text{ref}}_{\mu\nu}(Y + \sqrt{\alpha'} \eta)  \rangle.
\end{equation}
Expanding ${h}^{\text{ref}}_{\mu\nu}(Y^\mu + \sqrt{\alpha'} \eta^\mu)$ in the region near the wall where $\abs{Y^D}\sim \sqrt{\apr}$ gives\footnote{The component ${h}^{\text{ref}}_{iD }(X^D,\vec{X})$ does not contribute since we are expanding about a flat diagonal metric. Here $ {h}^{\text{ref}}_{iD }(X^D,\vec{X})$ has no $|X^D|$ source term. In the notation of \eqref{eq:psi_def}, it has $p(\bullet)=1$.}
\begin{equation}
\label{eq:h_mu_nu_expanded}
\begin{aligned}
h^{\text{ref}}_{ij }(X^D,\vec{X})&=h_{ij}(0,\vec{X})+\partial_D h_{ij}(0,\vec{X}){|}X^D{|}\\
h^{\text{ref}}_{DD}(X^D,\vec{X})&= h_{DD}(0,\vec{X})+\partial_D h_{DD}(0,\vec{X})|X^D| \\
h^{\text{ref}}_{iD}(X^D,\vec{X})&=\text{sgn}(X^D)h_{iD}(0,\vec{X})+\partial_D h_{iD}(0,\vec{X})X^D,
\end{aligned}
\end{equation}
where $\partial_D h_{\mu \nu}(0,\vec{X})$ is defined in \eqref{eq:derivative_at_wall}. After Taylor expanding ${h}^{\text{ref}}_{\mu \nu }(X^D,\vec{X})$ \eqref{eq:h_mu_nu_expanded} about the point $\vec{Y}$ to $O(\sqrt{\alpha'})$,\footnote{We expand to $O(\sqrt{\alpha'})$ because we when later integrate in the near wall region, the integral $\int_{|Y^D| \sim \sqrt{\alpha'}}dY^D (\cdots )$ introduces another factor of $\sqrt{\alpha'}$ such that this term contributes at order $\alpha'$. } the expectation value yields
\begin{equation}\label{eq:tensor_u(Y^D)}
\left\langle \partial_a \eta^\mu \partial_b \eta^\nu {h}^{\text{ref}}_{\mu \nu}(Y+\sqrt{\alpha'}\eta) \right\rangle= \left\langle \partial_a \eta^D \partial_b \eta^D \right\rangle \left( h_{\mu\mu}(0,\vec{Y})+\partial_Dh_{\mu\mu}(0,\vec{Y}) u(Y^D)\right),
\end{equation}
where we have used \eqref{eq:u(Y^D)}. In \eqref{eq:tensor_u(Y^D)}, we have used that
\begin{equation}
    \langle \partial_a \eta^\mu \partial_b \eta^\nu  \rangle = \delta^{\mu \nu} \langle \partial_a \eta^D \partial_b \eta^D \rangle,
\end{equation}
since different $\eta^\mu$ fields are decoupled.

Now we wish to express the metric perturbations $h_{\mu \mu}$ and their derivatives \eqref{eq:tensor_u(Y^D)} in terms of $K$. To achieve this, we first express them in terms of ADM variables \cite{amowitt1962dynamics,thorne2000gravitation,wald2010general,Poisson:2009pwt}
\begin{equation}\label{eq:ADM_metric}
ds^2= (1+h_{DD})(dX^D)^2+(\delta_{ij}+h_{ij})(dX^i+N^i dX^D) (dX^j+N^j dX^D)+O(h^2).
\end{equation}
Here the lapse function to linear order is $N=1+\frac{1}{2}h_{DD}$, the shift vector is $N^i = h_{iD}$, and the induced spatial metric on a hypersurface is $\gamma_{ij}=\delta_{ij}+h_{ij}$. The extrinsic curvature tensor $K_{ij}$ is then given by
\begin{equation}\label{eq:K_ij}
    K_{ij}=-\frac{1}{2 N}\left(\partial_D h_{ij} -\nabla_i N_j-\nabla_j N_i\right).
\end{equation}
The overall minus sign in \eqref{eq:K_ij} is due to the fact that the outward-pointing normal $n_{\mu}$ is  $\partial_n = - \partial_D$. To linear order in $h$, $K_{ij}$ is given by
\begin{equation}
    K_{ij}=-\frac{1}{2 N}\left(\partial_D h_{ij} -\partial_i h_{jD}-\partial_j h_{iD}+O(h^2)\right).
\end{equation}
Using that the inverse of $(\delta_{ij}+h_{ij})$ is $\delta_{ij}-h_{ij}+O(h^2)$, the trace $K$ to $O(h)$ takes the following form\footnote{The terms $(h_{ii}(0,\vec{Y}) + h_{DD}(0,\vec{Y})) \langle\partial_a \eta \partial_b \eta\rangle$ don't contribute since $\langle\partial_a \eta \partial_b \eta\rangle$ are purely quadratically divergent in $\eps$, and can therefore be removed by local counterterms.}
\begin{equation}\label{eq:TraceK}
\begin{aligned}
    K&= -(\delta_{ij}-h_{ij}+O(h^2))\frac{1}{2 }\left( 1+\frac{1}{2}h_{DD}\right)(\partial_D h_{ij} -\partial_i h_{jD}-\partial_j h_{iD})\\
    &=-\frac{1}{2}\partial_D h_{ii}+\partial_i h_{iD}+O(h^2).
    \end{aligned}
\end{equation}
In terms of \eqref{eq:TraceK}, we arrive at the following expression for \eqref{eq:tensor_u(Y^D)}\footnote{Note that $\partial_D h_{ii} = \partial_D h_{ii} - 2 \partial_i h_{iD}+2 \partial_i h_{iD}=-2K+2 \partial_i h_{iD}$.}
\begin{equation}
    \begin{aligned}
\langle \partial_a \eta^\mu \partial_b \eta^\nu {h}^{\text{ref}}_{\mu \nu}(Y+\sqrt{\alpha'}\eta) \rangle&=u(Y^D) \langle \partial_a \eta^D \partial_b \eta^D\rangle \left[-2K(0,\vec{Y})+2\partial_i h_{iD}(0,\vec{Y}) +\partial_Dh_{DD}(0,\vec{Y})\right].
    \end{aligned}
\end{equation}
Therefore, the expectation value \eqref{eq:EV_V_h} we would like to evaluate is
\begin{equation}\label{eq:V_def_metric}
  \langle  \Hat{V}_{\text{def}} \rangle  = \left[-2K(0,\vec{Y})+\partial_Dh_{DD}(0,\vec{Y}) +2\partial_i h_{iD}(0,\vec{Y})\right] u(Y^D) \frac{1}{4\pi} \int d^2 z \sqrt{g}  g^{ab}  \langle \partial_a \eta^D \partial_b \eta^D \rangle.
\end{equation}

We will next present the details of computing \eqref{eq:V_def_metric}. 

\section{The Einstein boundary term}\label{sec:EinsteinBoundaryAction}
In subsection \ref{sec:I_wall}, we evaluate the half-space partition function \eqref{eq:Z_HL} and derive $I_{\text{wall}}$ by applying Tseytlin's off-shell prescriptions. We then add it to the boundary term from the bulk tree-level action \eqref{eq:I_TD_intro} in subsection \ref{sec:I_total} to obtain the Einstein boundary term \eqref{eq:I_bdy_intro} of the $\Gamma^2$ action. 

As explained in section \ref{sec:MOI}, the half-space partition function \eqref{eq:Z_HL} can be expressed as the sum of two terms
\begin{align} \label{eq:Z_HL_split}
 Z_{\text{HS}}=\frac{1}{2}Z_{\text{nz}} V_{Y^D} + Z_{\text{wall}}+ Z_{\text{bulk}} + O(\Hat{V}_{\text{def}}^2),
\end{align}
where $V_{Y^D}=\int d^D Y$ and
\begin{align}
Z_{\text{wall}} &:=-\frac{1}{2} Z_{\text{nz}} \int_{\abs{Y^D}\sim \sqrt{\apr}} \, d^D Y  \langle \Hat{V}_{\text{def}} \rangle,  \\  
Z_{\text{bulk}}&:=-\frac{1}{2}Z_{\text{nz}} \int_{\abs{Y^D}\gg \sqrt{\apr}} \,d^D Y  \langle \Hat{V}_{\text{def}}  \rangle. 
\end{align}

Let us illustrate how we evaluate $Z_{\text{wall}}$.

\subsection{$I_{\text{wall}}$}\label{sec:I_wall}
To obtain the target space boundary term we integrate \eqref{eq:V_def_metric} near the boundary and then apply Tseytlin's sphere prescription \cite{TSEYTLINMobiusInfinitySubtraction1988,Ahmadain:2022eso} to eliminate the logarithmic divergence. Let us see how this works. In this subsection, we follow the same steps in Section 4 of \cite{Ahmadain_Khan} while omitting some details.

We want to evaluate $Z_{\text{HS}}$ \eqref{eq:Z_HL} near the wall where ${\abs{Y^D}\sim \sqrt{\apr}}$
\begin{align}\label{eq:Z_near}
Z_{\text{wall}}&=-\frac{1}{2} Z_{\text{nz}} \int_{\abs{Y^D}\sim \sqrt{\apr}} d^D Y \langle \Hat{V}_{\text{def}} \rangle .
\end{align}

To evaluate $\Hat{V}_{\text{def}}$ \eqref{eq:V_def_metric}, we use equation A.3 in appendix A of \cite{Tseytlin:1990mv} that $\int_{S^2} d^2 z \sqrt{g}  g^{ab}  \langle \partial_a \eta^\mu \partial_b \eta^\nu \rangle =2\pi \delta^{\mu \nu}(\bar{\delta}-1)$ after accounting for the difference in conventions. Note that in \cite{Tseytlin:1990mv} the mode splitting was $X(z) = Y+\eta(z)$ in units $2 \pi \alpha'=1$ as opposed to our convention where the mode splitting is $X(z) = Y+\sqrt{\alpha'}\eta(z)$. As mentioned in \cite{Tseytlin:1990mv}, setting $\bar{\delta}$ = 0 is the renormalization prescription which gives us a covariant answer; any other value of $\bar{\delta}$ does not. $\bar{\delta}$ = 0 removes power law and mixed power-logarithmic UV divergences from the path integral. With this prescription, in \eqref{eq:V_def_metric}, we get \begin{equation}\label{eq:delta_bar}
 \frac{1}{4\pi} \int_{S^2} d^2 z \sqrt{g}  g^{ab}  \langle \partial_a \eta^\mu \partial_b \eta^\nu \rangle   = -\frac{1}{2 }\delta_{\mu \nu}\,.
\end{equation}
We also need the following result from \cite{Ahmadain_Khan}
\begin{equation}\label{eq:u(Y)_integral}
\lim_{Y_c \to \infty}\int_{-\frac{Y_c \sqrt{\alpha^{\prime}}}{2}}^{\frac{Y_c \sqrt{\alpha^{\prime}}}{2}} d Y^D u(Y^D) = \apr \left( \frac{Y_c^2}{4} + \frac{\Omega}{2} \right) = \apr \left(\frac{ Y_c^2}{4} -\frac{\ln \eps}{2}\right).
\end{equation}
Here $Y_c$ is a target space infrared cutoff. We have expanded near $Y_c \to \infty$ and only kept the leading and subleading terms. We also used that $\Omega = -\ln \eps$ \cite{Ahmadain_Khan}. 

Plugging \eqref{eq:delta_bar} and \eqref{eq:u(Y)_integral} into \eqref{eq:Z_near}, we get
\begin{equation}\label{eq:Z_wall}
Z_{\text{wall}} = \apr Z_{\text{nz}} \left(-\frac{1}{2} \right)\int d^{D-1} Y \cdot   \text{Bd}[h](0,\vec{Y}) \left( \frac{ Y_c^2}{4} -\frac{\ln \eps}{2} \right),
\end{equation}
where
\begin{equation}
\text{Bd}[h](0,\vec{Y}):=K(0,\vec{Y})-\frac{1}{2}\partial_Dh_{DD}(0,\vec{Y}) -\partial_i h_{iD}(0,\vec{Y}).
\end{equation}

Applying Tseytlin's off-shell sphere prescription \cite{TSEYTLINMobiusInfinitySubtraction1988,Ahmadain:2022tew} to remove the logarithmic divergence in \eqref{eq:Z_wall}, we finally obtain the GHY term in addition to two other terms
\begin{equation}\label{eq:I_wall_TrK}
I_{\text{wall}} = -\frac{\partial}{\partial \ln \eps}Z_{\text{wall}} 
=  -\frac{1}{4}\apr Z_{\text{nz}} {\int d^{D-1} Y \cdot \left[K(0,\vec{Y})-\frac{1}{2}\partial_Dh_{DD}(0,\vec{Y}) -\partial_i h_{iD}(0,\vec{Y}) \right] }.
\end{equation}

In appendix \ref{app:gauge_terms}, we show that the last two terms to linear order in \eqref{eq:I_wall_TrK} can be expressed as
\begin{equation}\label{eq:two_gauge_terms}
- \frac{1}{2}\partial_D h_{DD}+O(h^2) = \frac{1}{2} n^\alpha n^\beta n^\mu \partial_\alpha G_{\mu \beta}, \quad   -\partial_i h_{iD}+O(h^2) = n^\beta \gamma^{\alpha \mu} \partial_\alpha G_{\mu \beta},
\end{equation}
and that $-\gamma^{\alpha \beta} \partial_\alpha n_\beta$ vanishes for the type of flat boundaries we are considering in this work. 
\paragraph{}
Therefore, to \emph{linear order}, we obtain 
\begin{equation}\label{eq:I_wall}
I_{\text{wall}} = -\apr \frac{Z_{\text{nz}}}{4} \int d^{D-1} Y \cdot \left[K+\frac{1}{2} n^\alpha n^\beta n^\mu \partial_\alpha G_{\mu \beta} + n^\beta \gamma^{\alpha \mu} \partial_\alpha G_{\mu \beta}\right].
\end{equation}
 This is the main result of our work.

\subsection{$I_{\text{bulk}}$}\label{sec:I_total}
In the bulk, far from the wall $|Y^D| \gg \sqrt{\alpha'}$, we have
\begin{equation}\label{eq:h^ref_bulk}
   {h}^{\text{ref}}_{\mu \nu}(Y+\sqrt{\alpha'}\eta)= (\text{sgn}(Y^D))^{\delta_{\mu,D}+\delta_{\nu,D}} \left(h_{\mu \nu}(Y) +\sqrt{\alpha'}\partial_\alpha {h}^{\text{ref}}_{\mu \nu}(Y) \eta^\alpha +\frac{\alpha'}{2}\partial_\alpha \partial_\beta{h}^{\text{ref}}_{\mu \nu}(Y) \eta^\alpha \eta^\beta +O({{\alpha'}}^{3/2})\right).
\end{equation}
In \eqref{eq:h^ref_bulk}, we do no sum over $\mu,\nu$ indices. The lowest order in $\alpha'$ of $\langle  \Hat{V}_{\text{def}} \rangle_{\text{bulk}}$ which contains a logarithmic divergence is given by
\begin{equation}
    \begin{aligned}
       \langle  \Hat{V}_{\text{def}} \rangle_{\text{bulk}}&=\frac{\alpha'}{2}g^{ab}\langle \partial_a \eta^\mu \partial_b \eta^\nu \rangle\langle \eta^\alpha \eta^\beta \rangle \partial_\alpha \partial_\beta h^{\text{ref}}_{\mu \nu}.
    \end{aligned}
\end{equation}
This gives
\begin{equation} 
\begin{aligned}
    Z_{\text{bulk}} &=-\frac{1}{2}Z_{\text{nz}} \int_{|{Y^D}|\gg \sqrt{\apr}} \,d^D Y ~ \frac{\alpha'}{2}\left( \frac{1}{4\pi} \int d^2 z \sqrt{g}  g^{ab}  \langle \partial_a \eta^\mu \partial_b \eta^\nu \rangle\right)\langle \eta^\alpha \eta^\beta \rangle\partial_\alpha \partial_\beta h^{\text{ref}}_{\mu \nu}\\
    &=-\frac{\alpha'}{2}Z_{\text{nz}}  \left( \frac{1}{4\pi} \int d^2 z \sqrt{g}  g^{ab}  \langle \partial_a \eta^\mu \partial_b \eta^\nu \rangle\right)\langle \eta^\alpha \eta^\beta \rangle \int_{Y^D\gg \sqrt{\apr}} \,d^D Y ~ \partial_\alpha \partial_\beta h_{\mu \nu},
\end{aligned}
\end{equation}
where we used \eqref{eq:delta_bar} and $\langle\eta^\mu \eta^\nu\rangle = \delta^{\mu \nu} \langle\eta^D\eta^D\rangle = \delta^{\mu \nu}\left(-\frac{1}{2}\ln \eps\right)$ \cite{Ahmadain_Khan}
\begin{align}
    Z_{\text{bulk}} = \frac{1}{8}\alpha' Z_{\text{nz}} (-\ln \eps) \int_{Y^D\gg \sqrt{\alpha'}} d^D Y ~\partial^2 h_{\mu \mu}.
\end{align}
Applying Tseytlin's prescription, we get
\begin{equation}
    I_{\text{bulk}} = \frac{1}{8}\alpha' Z_{\text{nz}} \int_{Y^D\gg \sqrt{\alpha'}} d^D Y ~\partial^2 h_{\mu \mu} =  \frac{1}{8}\alpha' Z_{\text{nz}} \int d^{D-1} Y ~n_\nu \partial^\nu h_{\mu \mu}(0,\vec{Y}).
\end{equation}
Note that for our wall at $-X^D =0$, the normal 1-form is $n_\mu = (-1, 0 ,\cdots,0)$. Then we obtain
\begin{equation}\label{eq:I_bulk_sec4}
  I_{\text{bulk}}  = - \frac{1}{8}\alpha' Z_{\text{nz}} \int d^{D-1} Y ~\partial_D h_{\mu \mu}(0,\vec{Y}) . 
\end{equation}
Note that $I_{\text{bulk}}$ in \eqref{eq:I_bulk_sec4} is equal at linear order to
\begin{equation}
\label{eqn: I bulk we calculate}
I_{\text{bulk}}=-\frac{1}{4}\alpha^{\prime} Z_{\text{nz}} \left[\int_{ \mathcal{M}}\mathrm{d}^D Y\, \sqrt{G} e^{-2\Phi} R +\int_{\partial \mathcal{M}} \mathrm{d}^{D-1} Y \sqrt{\gamma}\, e^{-2\Phi}\left(K-\gamma^{\alpha \beta} \partial_\alpha n_\beta-\frac{1}{2} n^\mu n^\alpha n^\beta \partial_\mu G_{\alpha \beta}\right)\right].
\end{equation}
Adding \eqref{eqn: I bulk we calculate}, \eqref{eq:I_wall}, and including the dilaton \eqref{I_dilaton_action}, we finally arrive at the off-shell classical effective action in half-space to $O(\apr)$
\begin{align}\label{eq:I_total_ch4}
I_{\text{sphere,HS}} &= -\frac{1}{4}\alpha' Z_{\text{nz}} \int_{\mathcal{M}} \mathrm{d}^D Y\, \sqrt{G}\, e^{-2 \Phi} (R + 2\nabla^2 \Phi) \nonumber \\
&\quad  - \frac{1}{4}\alpha' Z_{\text{nz}} \int_{\partial \mathcal{M}} \mathrm{d}^{D-1} Y \sqrt{\gamma}\, e^{-2 \Phi} \left(2 K - 2 \partial_n \Phi + n^\beta \gamma^{\alpha \mu} \partial_\alpha G_{\mu \beta}\right).
\end{align}


\section{Discussion and Outlook} \label{sec:discussion}

We have developed a novel approach using the method of images to derive target space boundary terms from the worldsheet NLSM. We have used it to study the motion of closed massless bosonic strings on a half-space and derive the total sphere effective action to $O(\apr)$ and to linear order in target spacetime metric and dilaton perturbation that has a well-posed variational principle. Using the dilaton equation of motion, we are then able to obtain the on-shell classical action for the metric and dilaton. The approach is quite general however. The methods can be extended to non-trivial string solutions, e.g. black hole backgrounds such as the cigar or BTZ, where the on-shell action directly yields the black hole entropy.

Now we would like to present a subtlety in our work. As we pointed out in section \ref{sec:intro}, the classical effective action we derive \eqref{eq:I_total} for closed strings in a half-space tells us that either Dirichlet \eqref{eq:DBC} or special Neumann boundary condition \eqref{eq:BY_Tensor_zero} should be imposed on the wall in order to have a well-defined variational principle. But we don't directly calculate $Z_{\text{HS}}$; rather, what we compute it using the method of images, and this is where the core of the problem lies.

The core of the puzzle lies in interpreting, within the context of string theory, what the string partition function calculated in the doubled space—where there are no explicit boundaries—corresponds to in the original half-space where a specific type of boundary condition is imposed at the wall. The challenge is to translate results from the boundary-less doubled space back to the half-space with explicit boundary conditions, ensuring that the physical quantities accurately reflect the constraints imposed at the wall. This involves understanding how the method of images encodes the boundary conditions into the symmetry of the extended space and how calculations there can be interpreted to yield meaningful results in the original domain with the boundary. 

Let us be more precise. Consider the motion of particle on a half-space in $D$-dimensional Euclidean half-space with coordinates $Y^{\mu}(\tau)$. The worldline of the particle is parametrized by Euclidean time $\tau$. In this case, the method of images provides a clear way to impose boundary conditions and compute quantities like the trace of the heat kernel. One way to compute $K(Y_1, Y_2; \beta)$ is through the path integral formulation, which sums over all possible paths $Y(\tau)$ from $Y_1$ to $Y_2$ in Euclidean time $\beta$
\be\label{eq:HK_Path_integral}
K(Y_1, Y_2; \beta) = \int_{Y(0)=Y_1}^{Y(\beta)=Y_2} [D Y] \, e^{-I[Y]}.
\ee
Here,  $[D Y] = \prod_{0<\tau<\beta} d^D Y(\tau)$ is the path integral measure over all paths in $\mathbb{R}^D$  for any $\tau$ and $I[Y]$ is the action along the path $Y(\tau)$.

The trace of the heat kernel is a \textit{one-loop} quantity represented by the path integral over closed paths with periodic boundary conditions\footnote{$H$ is the Hamiltonian of a nonrelativistic particle in a potential $V(Y)$ with mass $m=1$, $H=-\frac{1}{2} \nabla_Y^2+V(Y)$. The heat kernel $K(Y_1, Y_2 ; \beta)$ is the solution of the heat equation $-\frac{\partial}{\partial \beta} K(Y_1, Y_2 ; \beta)=H K(Y_1, Y_2 ; \beta)$ with the boundary condition
$K(Y_1, Y_2 ; 0)=\delta^D(Y_2 - Y_1)$.}

\be\label{eq:TraceHK}
\operatorname{Tr} e^{-\beta H} = \int_{\mathcal{M}} d^D Y \, K(Y, Y; \beta) = \int_{\text{PBC}} [D Y] \, e^{-I[Y]}.
\ee


In the presence of a boundary, such as in a half-space, the heat kernel in $\mathbb{R}_{+} \times \mathbb{R}^{D-1}$ can be computed using the method of images from the heat kernel in $\mathbb{R^D}$
\be\label{eq:HK_HL}
K_{\mathbb{R}_{+} \times \mathbb{R}^{D-1}}(Y_1, Y_2; \beta) = K_{\mathbb{R}^D}(Y_1, Y_2; \beta) \mp K_{\mathbb{R}^D}(Y_1, \tilde{Y}_2; \beta),
\ee
where $\tilde{Y}_2$ is the reflection of $Y_2$ across the boundary. The minus sign in \eqref{eq:HK_HL} corresponds to Dirichlet boundary conditions while the plus sign corresponds to Neumann boundary conditions. By adding or subtracting the heat kernel in $\mathbb{R^D}$ , we impose the desired boundary conditions at the wall on the \textit{wave function} evolved by heat kernel in the half-space $\mathbb{R}_{+} \times \mathbb{R}^{D-1}$ \cite{Bastianelli:2006hq,Bastianelli:2008vh}.

However, when we transition to the string worldsheet case, the situation becomes more intricate. The worldsheet path integral involves summing over two-dimensional surfaces with a given \textit{topology}. In string theory, the trace in \eqref{eq:TraceHK} corresponds to the \textit{annulus} partition function not the sphere partition function, which is the quantity we compute in our work. The annulus describes an open string of length $\pi$ propagating along a circle of length $2\pi \beta$ on each of the two boundaries of the annulus. In this description, the trace results from the identification of the initial and final states of the open string.\footnote{In the closed string channel, there is no trace; the annulus (cylinder) describes a closed string of length $2\pi\beta$ propagating between two ``branes" where they are emitted and then absorbed.} Since a sphere with two holes is homeomorphic to an annulus, one might consider cutting two discs on the sphere and imposing boundary conditions along their boundaries. We leave this for future work.

The moral is that, just like in the particle case, where the choice of bounce corresponds to selecting a boundary condition for the heat kernel, we conclude that the choice of bounce of the worldsheet on the wall also imposes a boundary condition in this case. This must be true because, once we choose the sign of the bounce of the worldsheet, the partition function is defined once and for all. Therefore, the choice of the sign of the bounce is the only place to impose any boundary conditions. However, unlike the particle case, where we understand which sign imposes which boundary conditions, we do not yet have a first-principles understanding of which sign imposes which boundary condition in the string case. One might naively think that it must be the same as in the particle case, but the quantities being computed in the two cases are different\footnote{We thank Arkadi Tseytlin for pointing this out to us.}. In the particle case, what is being computed is the one-loop correction to the quantum effective action, whereas in the string case, we compute the classical action. And according to our calculations, we find out that a $(+)$ sign gives the sphere action consistent with Dirichlet boundary conditions. We seek a deeper understanding of the stringy mechanism that determines the sign of the sphere partition function as the string bounces off the wall. It would be good to understand this at a more fundamental level.

In the above context, orbifolds with \textit{discrete torsion} \cite{Vafa:1986wx,Vafa:1994rv} may provide valuable insights into this important question\footnote{We thank Eva Silverstein for suggesting this analogy.}. This is despite the fact that our work is conceptually different from the motion of strings in orbifolds. It is not clear how boundaries are even defined in orbifolds and hence there is no notion of boundary conditions. Also, the singularity at the wall in this work is fundamentally different from the orbifold counterpart. However, the phase assigned to the twisted sector of the orbifold torus partition function, at least in spirit, looks similar to the the sign of the string's bounce at the wall, although discrte torsion only applies to the $B$-field.

\subsection{Outlook and Future Directions}
It should be relatively straightforward to derive the boundary term for the $B$-field and check if obtaining the correct coefficient would also require the addition of two contributions similar to the case of the metric and dilaton.


In this work as well as in \cite{Ahmadain_Khan}, we have deformed around flat half-space. An important generalization of this work is to apply the method of images to derive the on-shell action for a black hole string solution, e.g. the cigar \cite{Witten-BH:1991,Mandal:1991,Giveon:1991sy,Gibbons:1992rh} from which one can directly obtain the tree-level (classical) thermal entropy \cite{KT:2001,KKK:2000,Chen-LargeD:2021,chen2023black}, at least to first order in $\apr$. The cigar geometry is a semi-infinite cylinder that smoothly caps off at the tip. It would be exciting to apply the worldsheet approach developed here to describe the physics near the tip of the cigar and gain deeper insights into the nature of the tachyonic winding mode that dominates the physics in that region.

Another interesting direction is is to attempt a worldsheet derivation of the action for codimension-2 boundaries, such as the gravitational (Hayward) corner term for codimension-2 non-smooth boundaries \cite{Hayward:1993my,Jubb:2016qzt} which is arguably related to gravitational and stringy black hole edge modes \cite{Takayanagi-EE-StingTheory-2015}. In this case, one would have two walls located at $X^D=0$ and $X^{D-1} =0$ with a corner term at $X^D=X^{D-1} =0$ and four different regions such that the deformation looks something like
\begin{align}
\psi_{\operatorname{ref}}(X)&=  \psi\left(X^D, X^{D-1}, \vec{X}\right) \Theta\left(X^D\right) \Theta\left(X^{D-1}\right) \quad \text { (Region I) } \\
& +\mathcal{R}_{D-1} \circ \psi\left(X^D, X^{D-1}, \vec{X}\right) \Theta\left(X^D\right) \Theta\left(-X^{D-1}\right) \quad \text { (Region II) } \\
& +\mathcal{R}_D \circ \psi\left(X^D, X^{D-1}, \vec{X}\right) \Theta\left(-X^D\right) \Theta\left(X^{D-1}\right) \quad \text { (Region III) } \\
& +\mathcal{R}_{D,D-1} \circ \psi\left(X^D, X^{D-1}, \vec{X}\right) \Theta\left(-X^D\right) \Theta\left(-X^{D-1}\right) \quad \text { (Region IV) }.
\end{align}

It would be even more interesting to generalize the current work to strings propagating on orbifolds \cite{Dixon:1986qv} especially complex plane orbifold $\mathbb{C}/\mathbb{Z}_N$ which has been used to study black hole entropy in string theory \cite{Dabholkar-Orbifold1994,StromingerLowe-Orbifold-1994,Dabholkar-TachyonCond2002,Dabholkar:EntaglementStringTheory2022,Dabholkar:2023tzd}. The space can be visualized as a cone with a conical singularity at the orbifold fixed point and deficit angle $\Delta=2 \pi(1-\frac{1}{N})$. The $\mathbb{C}/\mathbb{Z}_N \times \mathbb{R}^{D-2}$ is an on-shell string background.

In this work, we used step functions to express deformations corresponding to reflections in a $\mathbb{R}/\mathbb{Z}_2$ orbifold. However, for $\mathbb{Z}_N$, which involves rotations, it's less straightforward what the appropriate deformations are. 


It would also be interesting to explore whether the method of images can be applied to derive the boundary term for conformal boundary conditions \cite{Anderson:2010ph,An:2021fcq} which have a well-posed initial value problem, in contrast to Dirichlet \cite{Anderson:2010ph,Witten:2018lgb}. In this case, rather than fixing the entire boundary metric, as in Dirichlet boundary conditions \eqref{eq:DBC}, we fix it up to a conformal factor together with $K$
\begin{equation}
\delta\left(\gamma^{-1 / (D-1)} \gamma_{\mu \nu}\right)=0, \quad \delta K=0 .
\end{equation}

There has been many recent exciting work on finite timelike boundaries in T$\bar{\text{T}}$-deformed theories \cite{Zamolodchikov:2004ce,Smirnov:2016lqw,Cavaglia:2016oda,McGough:2016lol,Taylor:2018xcy,Hartman:2018tkw,Gorbenko:2018oov,Lewkowycz:2019xse,Coleman:2020jte,Anninos:2023epi,Anninos:2024wpy,Silverstein:2024xnr}. An intriguing yet ambitious avenue is to explore the behavior of strings near these finite conformal and Dirichlet boundaries on a bulk Cauchy slice in \(\operatorname{AdS}\) \cite{Freidel:2008sh,Araujo-Regado:2022gvw}. An attempt was made in \cite{silverstein2023black} to construct examples in string/M-theory for bounding walls (on a cylinder) in an AdS black hole and in the dS static patch. Further probing of the behavior of strings near these totally absorbing walls is an exciting future direction. It would be particularly interesting if the methods developed in this paper could be generalized to explore the existence of these stringy D-walls.


\section*{Acknowledgements}
We are grateful to Aron Wall, Eva Silverstein, Prahar Mitra, Alex Frenkel, Raghu Mahajan, Arkadi Tseytlin, Vasudev Shyam, Zihan Yan, Minjae Cho, Xi Yin, Douglas Stanford, Edward Witten, David Kolchmeyer and Daniel Thompson for insightful discussions. We are especially grateful to Eva Silverstein and Aron Wall for giving valuable comments. AA is supported by The Royal Society and by STFC Consolidated Grant No. ST/X000648/1. S.A. is supported by National Science Foundation under grant No.2111998 and the Simons Foundation.
\paragraph{}
{\footnotesize {\bf Open Access Statement} - For the purpose of open access, the authors have applied a Creative Commons Attribution (CC BY) licence to any Author Accepted Manuscript version arising.}

\footnotesize{\textbf{Data access statement}: no new data were generated for this work.}

\begin{appendices}
\section{The gauge-dependent terms in the Einstein boundary action}\label{app:gauge_terms}
In this appendix, we show that the last two terms to linear order in \eqref{eq:I_wall_TrK} can be expressed as
\begin{equation}\label{eq:two_gauge_terms}
- \frac{1}{2}\partial_D h_{DD}+O(h^2) = \frac{1}{2} n^\alpha n^\beta n^\mu \partial_\alpha G_{\mu \beta}, \quad   -\partial_i h_{iD}+O(h^2) = n^\beta \gamma^{\alpha \mu} \partial_\alpha G_{\mu \beta}.
\end{equation}

We start by finding the normal vector to the surface defined by \( -X^D = 0 \). The normal 1-form \( n_\mu \) to the surface \( -X^D = 0 \) is given by the gradient of the defining function of the surface. Since the surface is defined by ``\(- X^D = 0 \)", the defining function \( f(X^\mu) \) is simply \( f(X^\mu) = -X^D \). Therefore, in this gauge choice, the normal 1-form is simply constant
\begin{equation}\label{eq:normal}
n_\mu = (-1,0, 0, \dots, 0).
\end{equation}
Note that the minus sign is important for the normal to point outward, in our particular case the normal is along negative $X^D$ axis.

Contracting with the inverse metric $G^{\mu\nu}$ gives the normal vector $n^\mu$ 
\begin{equation}\label{eq:Inverse_normal}
n^\mu = G^{\mu\nu} n_\nu = -G^{\mu D}.
\end{equation}

The target space metric $G_{\alpha \beta}$ is the sum of the spatial and normal components
\begin{equation}
G_{\alpha \beta} = \gamma_{\alpha \beta}+n_\alpha n_\beta.
\end{equation}

In terms of the metric \eqref{eq:ADM_metric}, which we reproduce here for convenience
\begin{equation}\label{eq:ADM_metric_app}
ds^2= (1+h_{DD})(dX^D)^2+(\delta_{ij}+h_{ij})(dX^i+N^i dX^D) (dX^j+N^j dX^D)+O(h^2),
\end{equation}
we see that
\begin{equation}
\gamma_{ij}=\delta_{ij}+h_{ij}, \quad \gamma_{iD}=h_{iD}, \quad \gamma_{DD}=h_{DD},
\end{equation}
where $\gamma_{ij}$ is the induced metric on a spatial hypersurface with indices $i,j \in \{1,D-1\}$.

The inverse metric $G^{\alpha \beta}$ in terms of $h_{\alpha \beta}$ is given by to $O(h)$
\begin{equation}\label{eq:InverseMetric}
G^{\alpha \beta}= \delta^{\alpha \beta}-h^{\alpha\beta} +O(h^2).
\end{equation}

Using \eqref{eq:ADM_metric_app} and \eqref{eq:InverseMetric}, $n^\mu$ \eqref{eq:Inverse_normal} to $O(h)$ is given by 
\begin{equation}
n^\mu = (n^D,n^i)  = \left(-1 + h_{DD},h_{iD}\right).
\end{equation}

Similarly, using \eqref{eq:InverseMetric}, we find that to $O(h)$, the components of $\gamma^{\mu \nu}= G^{\mu \alpha}G^{\nu \beta}\gamma_{\alpha \beta}$ are given by
\begin{align}
\gamma^{ij}= \delta_{ij}-h_{ij}, \quad \gamma^{iD}=0, \quad \gamma^{DD}=h_{DD}.
\end{align}

There are three gauge-dependent terms in \eqref{eq:I_wall}. The term
\begin{equation}\label{eq:First_Gauge_term}
-\gamma^{\alpha \beta} \partial_\alpha n_\beta,
\end{equation}
represents the divergence of the normal $n_{\alpha}$ projected onto the boundary hypersurface.  This term is nonzero when the normal 1-form $n_\nu$ varies along the hypersurface, which occurs if it is curved within the ambient space. In our case, the wall defined by $X^D=0$ has a flat embedding in the ambient space, and the normal 1-form $n_\nu$ \eqref{eq:normal} is constant across the hypersurface. As a result, the derivative $\partial_\mu n_\nu$ is zero, and therefore the term $h^{\mu \nu} \partial_\mu n_\nu$ \eqref{eq:First_Gauge_term} vanishes. Note that this corresponds to a specific coordinate choice (or gauge) where the hypersurface is orthogonal to the coordinate lines.

Our calculation of $I_{\text{wall}}$ produces however the other two terms \eqref{eq:two_gauge_terms} in the Einstein boundary action. The first term in \eqref{eq:two_gauge_terms} is 
\begin{align}
    \frac{1}{2} n^\alpha n^\beta n^\mu \partial_\alpha G_{\mu \beta}.
\end{align}
Since $\partial_\alpha G_{\mu \beta}$ is $O(h)$, $n^\alpha n^\beta n^\mu$ only contributes to zeroth order. This is only possible when $\alpha = \beta =\mu = D$. Using \eqref{eq:Inverse_normal}, we therfore find
\begin{align}
\frac{1}{2} n^\alpha n^\beta n^\mu \partial_\alpha G_{\mu \beta}= \frac{1}{2}(-1)^3 \partial_D h_{DD}=- \frac{1}{2}\partial_D h_{DD}.
\end{align}

Now we move to the second term in \eqref{eq:two_gauge_terms}
\begin{align}
n^\beta \gamma^{\alpha \mu} \partial_\alpha G_{\mu \beta}.
\end{align}
As in the first term, since $\partial_\alpha G_{\mu \beta}$ is $O(h)$,  $n^\beta \gamma^{\alpha \mu}$ contributes only to zeroth order. Setting $\beta=D$ and using \eqref{eq:Inverse_normal}, this becomes
\begin{align}
n^\beta \gamma^{\alpha \mu} \partial_\alpha G_{\mu \beta} =  -\gamma^{\alpha \mu} \partial_\alpha G_{\mu D},
\end{align}
which can be non-zero only when $\alpha=i$ and $\mu=j$, so we get
\begin{align}
n^\beta \gamma^{\alpha \mu} \partial_\alpha G_{\mu \beta} =  -\gamma^{\alpha \mu} \partial_\alpha G_{\mu D}= -\gamma^{ij} \partial_i h_{j D}=-(\delta_{ij}-h_{ij})\partial_i h_{j D}=-\partial_i h_{iD}.
\end{align}

This proves \eqref{eq:two_gauge_terms}.

\end{appendices}

\bibliographystyle{JHEP}
\bibliography{main.bib}

\end{document}